\pdfoutput=1
\documentclass[11pt,twoside,a4paper,cmspaper,final,collab]{cms-tdr}
\def\svnVersion{6954}\def\cmsCernNo{2010-011}\def\cmsCernDate{\today}

\begin{document}\cmsNoteHeader{MUO-10-001}
%
%
%

%
%
\hyphenation{env-iron-men-tal}
\hyphenation{had-ron-i-za-tion}
\hyphenation{cal-or-i-me-ter}
\hyphenation{de-vices}
%
\RCS$Revision: 6857 $
\RCS$HeadURL: svn+ssh://alverson@svn.cern.ch/reps/tdr2/papers/MUO-10-001/trunk/MUO-10-001.tex $
\RCS$Id: MUO-10-001.tex 6857 2010-05-27 10:43:57Z pablog $
%
%
%

\providecommand {\etal}{\mbox{et al.}\xspace} 
\providecommand {\ie}{\mbox{i.e.}\xspace}     
\providecommand {\eg}{\mbox{e.g.}\xspace}     
\providecommand {\etc}{\mbox{etc.}\xspace}     
\providecommand {\vs}{\mbox{\sl vs.}\xspace}      
\providecommand {\mdash}{\ensuremath{\mathrm{-}}} 

\providecommand {\Lone}{Level-1\xspace} 
\providecommand {\Ltwo}{Level-2\xspace}
\providecommand {\Lthree}{Level-3\xspace}

\providecommand{\ACERMC} {\textsc{AcerMC}\xspace}
\providecommand{\ALPGEN} {{\textsc{alpgen}}\xspace}
\providecommand{\CHARYBDIS} {{\textsc{charybdis}}\xspace}
\providecommand{\CMKIN} {\textsc{cmkin}\xspace}
\providecommand{\CMSIM} {{\textsc{cmsim}}\xspace}
\providecommand{\CMSSW} {{\textsc{cmssw}}\xspace}
\providecommand{\COBRA} {{\textsc{cobra}}\xspace}
\providecommand{\COCOA} {{\textsc{cocoa}}\xspace}
\providecommand{\COMPHEP} {\textsc{CompHEP}\xspace}
\providecommand{\EVTGEN} {{\textsc{evtgen}}\xspace}
\providecommand{\FAMOS} {{\textsc{famos}}\xspace}
\providecommand{\GARCON} {\textsc{garcon}\xspace}
\providecommand{\GARFIELD} {{\textsc{garfield}}\xspace}
\providecommand{\GEANE} {{\textsc{geane}}\xspace}
\providecommand{\GEANTfour} {{\textsc{geant4}}\xspace}
\providecommand{\GEANTthree} {{\textsc{geant3}}\xspace}
\providecommand{\GEANT} {{\textsc{geant}}\xspace}
\providecommand{\HDECAY} {\textsc{hdecay}\xspace}
\providecommand{\HERWIG} {{\textsc{herwig}}\xspace}
\providecommand{\HIGLU} {{\textsc{higlu}}\xspace}
\providecommand{\HIJING} {{\textsc{hijing}}\xspace}
\providecommand{\IGUANA} {\textsc{iguana}\xspace}
\providecommand{\ISAJET} {{\textsc{isajet}}\xspace}
\providecommand{\ISAPYTHIA} {{\textsc{isapythia}}\xspace}
\providecommand{\ISASUGRA} {{\textsc{isasugra}}\xspace}
\providecommand{\ISASUSY} {{\textsc{isasusy}}\xspace}
\providecommand{\ISAWIG} {{\textsc{isawig}}\xspace}
\providecommand{\MADGRAPH} {\textsc{MadGraph}\xspace}
\providecommand{\MCATNLO} {\textsc{mc@nlo}\xspace}
\providecommand{\MCFM} {\textsc{mcfm}\xspace}
\providecommand{\MILLEPEDE} {{\textsc{millepede}}\xspace}
\providecommand{\ORCA} {{\textsc{orca}}\xspace}
\providecommand{\OSCAR} {{\textsc{oscar}}\xspace}
\providecommand{\PHOTOS} {\textsc{photos}\xspace}
\providecommand{\PROSPINO} {\textsc{prospino}\xspace}
\providecommand{\PYTHIA} {{\textsc{pythia}}\xspace}
\providecommand{\SHERPA} {{\textsc{sherpa}}\xspace}
\providecommand{\TAUOLA} {\textsc{tauola}\xspace}
\providecommand{\TOPREX} {\textsc{TopReX}\xspace}
\providecommand{\XDAQ} {{\textsc{xdaq}}\xspace}

\providecommand {\DZERO}{D\O\xspace}     


\providecommand{\de}{\ensuremath{^\circ}}
\providecommand{\ten}[1]{\ensuremath{\times \text{10}^\text{#1}}}
\providecommand{\unit}[1]{\ensuremath{\text{\,#1}}\xspace}
\providecommand{\mum}{\ensuremath{\,\mu\text{m}}\xspace}
\providecommand{\micron}{\ensuremath{\,\mu\text{m}}\xspace}
\providecommand{\cm}{\ensuremath{\,\text{cm}}\xspace}
\providecommand{\mm}{\ensuremath{\,\text{mm}}\xspace}
\providecommand{\mus}{\ensuremath{\,\mu\text{s}}\xspace}
\providecommand{\keV}{\ensuremath{\,\text{ke\hspace{-.08em}V}}\xspace}
\providecommand{\MeV}{\ensuremath{\,\text{Me\hspace{-.08em}V}}\xspace}
\providecommand{\GeV}{\ensuremath{\,\text{Ge\hspace{-.08em}V}}\xspace}
\providecommand{\TeV}{\ensuremath{\,\text{Te\hspace{-.08em}V}}\xspace}
\providecommand{\PeV}{\ensuremath{\,\text{Pe\hspace{-.08em}V}}\xspace}
\providecommand{\keVc}{\ensuremath{{\,\text{ke\hspace{-.08em}V\hspace{-0.16em}/\hspace{-0.08em}}c}}\xspace}
\providecommand{\MeVc}{\ensuremath{{\,\text{Me\hspace{-.08em}V\hspace{-0.16em}/\hspace{-0.08em}}c}}\xspace}
\providecommand{\GeVc}{\ensuremath{{\,\text{Ge\hspace{-.08em}V\hspace{-0.16em}/\hspace{-0.08em}}c}}\xspace}
\providecommand{\TeVc}{\ensuremath{{\,\text{Te\hspace{-.08em}V\hspace{-0.16em}/\hspace{-0.08em}}c}}\xspace}
\providecommand{\keVcc}{\ensuremath{{\,\text{ke\hspace{-.08em}V\hspace{-0.16em}/\hspace{-0.08em}}c^\text{2}}}\xspace}
\providecommand{\MeVcc}{\ensuremath{{\,\text{Me\hspace{-.08em}V\hspace{-0.16em}/\hspace{-0.08em}}c^\text{2}}}\xspace}
\providecommand{\GeVcc}{\ensuremath{{\,\text{Ge\hspace{-.08em}V\hspace{-0.16em}/\hspace{-0.08em}}c^\text{2}}}\xspace}
\providecommand{\TeVcc}{\ensuremath{{\,\text{Te\hspace{-.08em}V\hspace{-0.16em}/\hspace{-0.08em}}c^\text{2}}}\xspace}

\providecommand{\pbinv} {\mbox{\ensuremath{\,\text{pb}^\text{$-$1}}}\xspace}
\providecommand{\fbinv} {\mbox{\ensuremath{\,\text{fb}^\text{$-$1}}}\xspace}
\providecommand{\nbinv} {\mbox{\ensuremath{\,\text{nb}^\text{$-$1}}}\xspace}
\providecommand{\percms}{\ensuremath{\,\text{cm}^\text{$-$2}\,\text{s}^\text{$-$1}}\xspace}
\providecommand{\lumi}{\ensuremath{\mathcal{L}}\xspace}
\providecommand{\Lumi}{\ensuremath{\mathcal{L}}\xspace}
%
\providecommand{\LvLow}  {\ensuremath{\mathcal{L}=\text{10}^\text{32}\,\text{cm}^\text{$-$2}\,\text{s}^\text{$-$1}}\xspace}
\providecommand{\LLow}   {\ensuremath{\mathcal{L}=\text{10}^\text{33}\,\text{cm}^\text{$-$2}\,\text{s}^\text{$-$1}}\xspace}
\providecommand{\lowlumi}{\ensuremath{\mathcal{L}=\text{2}\times \text{10}^\text{33}\,\text{cm}^\text{$-$2}\,\text{s}^\text{$-$1}}\xspace}
\providecommand{\LMed}   {\ensuremath{\mathcal{L}=\text{2}\times \text{10}^\text{33}\,\text{cm}^\text{$-$2}\,\text{s}^\text{$-$1}}\xspace}
\providecommand{\LHigh}  {\ensuremath{\mathcal{L}=\text{10}^\text{34}\,\text{cm}^\text{-2}\,\text{s}^\text{$-$1}}\xspace}
\providecommand{\hilumi} {\ensuremath{\mathcal{L}=\text{10}^\text{34}\,\text{cm}^\text{-2}\,\text{s}^\text{$-$1}}\xspace}


\providecommand{\zp}{\ensuremath{\mathrm{Z}^\prime}\xspace}


\providecommand{\kt}{\ensuremath{k_{\mathrm{T}}}\xspace}
\providecommand{\BC}{\ensuremath{{B_{\mathrm{c}}}}\xspace}
\providecommand{\bbarc}{\ensuremath{{\overline{\mathrm{b}}\mathrm{c}}}\xspace}
\providecommand{\bbbar}{\ensuremath{{\mathrm{b}\overline{\mathrm{b}}}}\xspace}
\providecommand{\ccbar}{\ensuremath{{\mathrm{c}\overline{\mathrm{c}}}}\xspace}
\providecommand{\JPsi}{\ensuremath{{\mathrm{J}}\hspace{-.08em}/\hspace{-.14em}\psi}\xspace}
\providecommand{\bspsiphi}{\ensuremath{\mathrm{B}_\mathrm{s} \to \JPsi\, \phi}\xspace}
\providecommand{\AFB}{\ensuremath{A_\text{FB}}\xspace}
\providecommand{\EE}{\ensuremath{\mathrm{e}^+\mathrm{e}^-}\xspace}
\providecommand{\MM}{\ensuremath{\mu^+\mu^-}\xspace}
\providecommand{\TT}{\ensuremath{\tau^+\tau^-}\xspace}
\providecommand{\wangle}{\ensuremath{\sin^{2}\theta_{\text{eff}}^\text{lept}(M^2_\mathrm{Z})}\xspace}
\providecommand{\ttbar}{\ensuremath{{\mathrm{t}\overline{\mathrm{t}}}}\xspace}
\providecommand{\stat}{\ensuremath{\,\text{(stat.)}}\xspace}
\providecommand{\syst}{\ensuremath{\,\text{(syst.)}}\xspace}

\providecommand{\HGG}{\ensuremath{\mathrm{H}\to\gamma\gamma}}
\providecommand{\gev}{\GeV}
\providecommand{\GAMJET}{\ensuremath{\gamma + \text{jet}}}
\providecommand{\PPTOJETS}{\ensuremath{\mathrm{pp}\to\text{jets}}}
\providecommand{\PPTOGG}{\ensuremath{\mathrm{pp}\to\gamma\gamma}}
\providecommand{\PPTOGAMJET}{\ensuremath{\mathrm{pp}\to\gamma + \mathrm{jet}}}
\providecommand{\MH}{\ensuremath{\mathrm{M_{\mathrm{H}}}}}
\providecommand{\RNINE}{\ensuremath{\mathrm{R}_\mathrm{9}}}
\providecommand{\DR}{\ensuremath{\Delta\mathrm{R}}}


\providecommand{\PT}{\ensuremath{p_{\mathrm{T}}}\xspace}
\providecommand{\pt}{\ensuremath{p_{\mathrm{T}}}\xspace}
\providecommand{\ET}{\ensuremath{E_{\mathrm{T}}}\xspace}
\providecommand{\HT}{\ensuremath{H_{\mathrm{T}}}\xspace}
\providecommand{\et}{\ensuremath{E_{\mathrm{T}}}\xspace}
\providecommand{\Em}{\ensuremath{E\!\!\!/}\xspace}
\providecommand{\Pm}{\ensuremath{p\!\!\!/}\xspace}
\providecommand{\PTm}{\ensuremath{{p\!\!\!/}_{\mathrm{T}}}\xspace}
\providecommand{\ETm}{\ensuremath{E_{\mathrm{T}}^{\text{miss}}}\xspace}
\providecommand{\MET}{\ensuremath{E_{\mathrm{T}}^{\text{miss}}}\xspace}
\providecommand{\ETmiss}{\ensuremath{E_{\mathrm{T}}^{\text{miss}}}\xspace}
\providecommand{\VEtmiss}{\ensuremath{{\vec E}_{\mathrm{T}}^{\text{miss}}}\xspace}

\providecommand{\dd}[2]{\ensuremath{\frac{\mathrm{d} #1}{\mathrm{d} #2}}}

%

\providecommand{\ga}{\ensuremath{\gtrsim}}
\providecommand{\la}{\ensuremath{\lesssim}}
\providecommand{\swsq}{\ensuremath{\sin^2\theta_\mathrm{W}}\xspace}
\providecommand{\cwsq}{\ensuremath{\cos^2\theta_\mathrm{W}}\xspace}
\providecommand{\tanb}{\ensuremath{\tan\beta}\xspace}
\providecommand{\tanbsq}{\ensuremath{\tan^{2}\beta}\xspace}
\providecommand{\sidb}{\ensuremath{\sin 2\beta}\xspace}
\providecommand{\alpS}{\ensuremath{\alpha_S}\xspace}
\providecommand{\alpt}{\ensuremath{\tilde{\alpha}}\xspace}

\providecommand{\QL}{\ensuremath{Q_L}\xspace}
\providecommand{\sQ}{\ensuremath{\tilde{Q}}\xspace}
\providecommand{\sQL}{\ensuremath{\tilde{Q}_L}\xspace}
\providecommand{\ULC}{\ensuremath{U_L^C}\xspace}
\providecommand{\sUC}{\ensuremath{\tilde{U}^C}\xspace}
\providecommand{\sULC}{\ensuremath{\tilde{U}_L^C}\xspace}
\providecommand{\DLC}{\ensuremath{D_L^C}\xspace}
\providecommand{\sDC}{\ensuremath{\tilde{D}^C}\xspace}
\providecommand{\sDLC}{\ensuremath{\tilde{D}_L^C}\xspace}
\providecommand{\LL}{\ensuremath{L_L}\xspace}
\providecommand{\sL}{\ensuremath{\tilde{L}}\xspace}
\providecommand{\sLL}{\ensuremath{\tilde{L}_L}\xspace}
\providecommand{\ELC}{\ensuremath{E_L^C}\xspace}
\providecommand{\sEC}{\ensuremath{\tilde{E}^C}\xspace}
\providecommand{\sELC}{\ensuremath{\tilde{E}_L^C}\xspace}
\providecommand{\sEL}{\ensuremath{\tilde{E}_L}\xspace}
\providecommand{\sER}{\ensuremath{\tilde{E}_R}\xspace}
\providecommand{\sFer}{\ensuremath{\tilde{f}}\xspace}
\providecommand{\sQua}{\ensuremath{\tilde{q}}\xspace}
\providecommand{\sUp}{\ensuremath{\tilde{u}}\xspace}
\providecommand{\suL}{\ensuremath{\tilde{u}_L}\xspace}
\providecommand{\suR}{\ensuremath{\tilde{u}_R}\xspace}
\providecommand{\sDw}{\ensuremath{\tilde{d}}\xspace}
\providecommand{\sdL}{\ensuremath{\tilde{d}_L}\xspace}
\providecommand{\sdR}{\ensuremath{\tilde{d}_R}\xspace}
\providecommand{\sTop}{\ensuremath{\tilde{t}}\xspace}
\providecommand{\stL}{\ensuremath{\tilde{t}_L}\xspace}
\providecommand{\stR}{\ensuremath{\tilde{t}_R}\xspace}
\providecommand{\stone}{\ensuremath{\tilde{t}_1}\xspace}
\providecommand{\sttwo}{\ensuremath{\tilde{t}_2}\xspace}
\providecommand{\sBot}{\ensuremath{\tilde{b}}\xspace}
\providecommand{\sbL}{\ensuremath{\tilde{b}_L}\xspace}
\providecommand{\sbR}{\ensuremath{\tilde{b}_R}\xspace}
\providecommand{\sbone}{\ensuremath{\tilde{b}_1}\xspace}
\providecommand{\sbtwo}{\ensuremath{\tilde{b}_2}\xspace}
\providecommand{\sLep}{\ensuremath{\tilde{l}}\xspace}
\providecommand{\sLepC}{\ensuremath{\tilde{l}^\mathrm{C}}\xspace}
\providecommand{\sEl}{\ensuremath{\tilde{\mathrm{e}}}\xspace}
\providecommand{\sElC}{\ensuremath{\tilde{\mathrm{e}}^\mathrm{C}}\xspace}
\providecommand{\seL}{\ensuremath{\tilde{\mathrm{e}}_\mathrm{L}}\xspace}
\providecommand{\seR}{\ensuremath{\tilde{\mathrm{e}}_\mathrm{R}}\xspace}
\providecommand{\snL}{\ensuremath{\tilde{\nu}_L}\xspace}
\providecommand{\sMu}{\ensuremath{\tilde{\mu}}\xspace}
\providecommand{\sNu}{\ensuremath{\tilde{\nu}}\xspace}
\providecommand{\sTau}{\ensuremath{\tilde{\tau}}\xspace}
\providecommand{\Glu}{\ensuremath{g}\xspace}
\providecommand{\sGlu}{\ensuremath{\tilde{g}}\xspace}
\providecommand{\Wpm}{\ensuremath{\mathrm{W}^{\pm}}\xspace}
\providecommand{\sWpm}{\ensuremath{\tilde{\mathrm{W}}^{\pm}}\xspace}
\providecommand{\Wz}{\ensuremath{\mathrm{W}^{0}}\xspace}
\providecommand{\sWz}{\ensuremath{\tilde{\mathrm{W}}^{0}}\xspace}
\providecommand{\sWino}{\ensuremath{\tilde{\mathrm{W}}}\xspace}
\providecommand{\Bz}{\ensuremath{\mathrm{B}^{0}}\xspace}
\providecommand{\sBz}{\ensuremath{\tilde{\mathrm{B}}^{0}}\xspace}
\providecommand{\sBino}{\ensuremath{\tilde{\mathrm{B}}}\xspace}
\providecommand{\Zz}{\ensuremath{\mathrm{Z}^{0}}\xspace}
\providecommand{\sZino}{\ensuremath{\tilde{\mathrm{Z}}^{0}}\xspace}
\providecommand{\sGam}{\ensuremath{\tilde{\gamma}}\xspace}
\providecommand{\chiz}{\ensuremath{\tilde{\chi}^{0}}\xspace}
\providecommand{\chip}{\ensuremath{\tilde{\chi}^{+}}\xspace}
\providecommand{\chim}{\ensuremath{\tilde{\chi}^{-}}\xspace}
\providecommand{\chipm}{\ensuremath{\tilde{\chi}^{\pm}}\xspace}
\providecommand{\Hone}{\ensuremath{\mathrm{H}_\mathrm{d}}\xspace}
\providecommand{\sHone}{\ensuremath{\tilde{\mathrm{H}}_\mathrm{d}}\xspace}
\providecommand{\Htwo}{\ensuremath{\mathrm{H}_\mathrm{u}}\xspace}
\providecommand{\sHtwo}{\ensuremath{\tilde{\mathrm{H}}_\mathrm{u}}\xspace}
\providecommand{\sHig}{\ensuremath{\tilde{\mathrm{H}}}\xspace}
\providecommand{\sHa}{\ensuremath{\tilde{\mathrm{H}}_\mathrm{a}}\xspace}
\providecommand{\sHb}{\ensuremath{\tilde{\mathrm{H}}_\mathrm{b}}\xspace}
\providecommand{\sHpm}{\ensuremath{\tilde{\mathrm{H}}^{\pm}}\xspace}
\providecommand{\hz}{\ensuremath{\mathrm{h}^{0}}\xspace}
\providecommand{\Hz}{\ensuremath{\mathrm{H}^{0}}\xspace}
\providecommand{\Az}{\ensuremath{\mathrm{A}^{0}}\xspace}
\providecommand{\Hpm}{\ensuremath{\mathrm{H}^{\pm}}\xspace}
\providecommand{\sGra}{\ensuremath{\tilde{\mathrm{G}}}\xspace}
\providecommand{\mtil}{\ensuremath{\tilde{m}}\xspace}
\providecommand{\rpv}{\ensuremath{\rlap{\kern.2em/}R}\xspace}
\providecommand{\LLE}{\ensuremath{LL\bar{E}}\xspace}
\providecommand{\LQD}{\ensuremath{LQ\bar{D}}\xspace}
\providecommand{\UDD}{\ensuremath{\overline{UDD}}\xspace}
\providecommand{\Lam}{\ensuremath{\lambda}\xspace}
\providecommand{\Lamp}{\ensuremath{\lambda'}\xspace}
\providecommand{\Lampp}{\ensuremath{\lambda''}\xspace}
\providecommand{\spinbd}[2]{\ensuremath{\bar{#1}_{\dot{#2}}}\xspace}

\providecommand{\MD}{\ensuremath{{M_\mathrm{D}}}\xspace}
\providecommand{\Mpl}{\ensuremath{{M_\mathrm{Pl}}}\xspace}
\providecommand{\Rinv} {\ensuremath{{R}^{-1}}\xspace}
\newcommand{\pho}{\phantom{0}}
\newcommand{\phoo}{\phantom{00}}

\cmsNoteHeader{MUO-10-001}
\title{Measurement of the charge ratio of atmospheric muons with the CMS detector}
\address[cern]{CERN}
\author[cern]{The CMS Collaboration}
 
\hypersetup{%
pdfauthor={CMS Collaboration},%
pdftitle={Measurement of the charge ratio of atmospheric muons with the CMS detector},%
pdfsubject={CMS},%
pdfkeywords={CMS, physics, muon, cosmic rays, charge ratio}}

\date{\today}

\abstract{
We present a measurement of the ratio of positive to negative muon fluxes from 
cosmic ray interactions in the atmosphere, using data collected by the CMS detector 
both at ground level and in the underground experimental cavern at the CERN LHC.
Muons were detected in the momentum range from 
5\ensuremath{{\,\text{Ge\hspace{-.08em}V\hspace{-0.16em}/\hspace{-0.08em}}c}}\xspace
to 1\ensuremath{{\,\text{Te\hspace{-.08em}V\hspace{-0.16em}/\hspace{-0.08em}}c}}\xspace.
The surface flux ratio is measured to be 
$1.2766 \pm 0.0032 \, \mathrm{(stat.)} \pm 0.0032 \, \mathrm{(syst.)}$, independent
of the muon momentum, below 100\ensuremath{{\,\text{Ge\hspace{-.08em}V\hspace{-0.16em}/\hspace{-0.08em}}c}}\xspace.
This is the most precise measurement to date. 
At higher momenta the data are consistent with an increase of the charge ratio, in 
agreement with cosmic ray shower models and compatible with previous measurements 
by deep-underground experiments.
}

\maketitle



\section{Introduction}

The muon charge ratio $R$ is defined as the ratio of the number of positive- to
negative-charge atmospheric muons arriving at the Earth's surface.
These muons arise from showers produced in
interactions of high-energy cosmic ray particles with air 
nuclei in the upper layers of the atmosphere.
The magnitude and the momentum dependence of $R$ are determined by the
production and interaction cross sections of mesons (mainly pions and kaons),
and by their decay lengths.
As most cosmic rays and the nuclei with which they interact are positively charged,
positive meson production is favoured, hence more positive muons are expected.
Previous measurements from various
experiments~\cite{utah,baxendale,rastin,hebbeker,L3C,Adamson:2007ww,arXiv09063726,OPERA}
showed the muon charge ratio to be constant up to a momentum of about
$200 \GeVc$, and then to increase at higher 
momenta, in agreement with the predicted rise in the fraction of muons from kaon decays.
Measurements of the charge ratio can be used to constrain 
hadronic interaction models and to predict better the atmospheric neutrino flux.

The Compact Muon Solenoid (CMS)~\cite{cms} is one of the detectors
installed at the Large Hadron Collider (LHC)~\cite{lhc} at CERN.
The main goal of the CMS experiment is to search for signals of new physics
in proton-proton collisions at
centre-of-mass energies from 7 to 14\TeV{}~\cite{ptdr}.

Cosmic rays were used extensively to commission the CMS detector~\cite{mtcc,CRAFT08}.
These data can also
be used to perform measurements of physical quantities related to cosmic ray muons.
This letter presents a measurement of the muon charge ratio using CMS data
collected in two cosmic ray runs in the years 2006 and 2008.
More details of the analyses can be found \mbox{in~\cite{CMS_NOTE_2008-016,MUO-10-001}}.

\section{Experimental setup, data samples, and event simulation}
\label{sec:experiment}

The central feature of the CMS apparatus is a superconducting solenoid, of 6~m
internal diameter, providing a field of 3.8~T. Within the field volume are the
silicon pixel and strip tracker~\cite{TOB}, the crystal electromagnetic calorimeter
and the brass-scintillator hadron calorimeter.
Muons are measured in gas-ionization detectors embedded in the steel return
yokes~\cite{muonTDR}.
In the barrel there is a Drift Tube (DT) system interspersed with Resistive
Plate Chambers (RPCs), and in the endcaps there is a Cathode Strip Chamber (CSC) system,
also interspersed with RPCs. In addition to the barrel and endcap
detectors, CMS has extensive forward calorimetry.
A detailed description of CMS can be found in~\cite{cms}.

The CMS detector is installed in an underground cavern, with the center 
of the detector 89~m below Earth's surface, and 420~m above sea level.
The location is $\,46^\circ\,18.57^\prime$ north latitude and
$\,6^\circ\,4.62^\prime$ east longitude.  The upper 50~m of the material
above CMS consists of moraines, followed by 20~m of molasse 
rock. A large access shaft with a diameter of 20.5~m rises vertically to the 
surface, and is offset from the center of CMS by 14~m along the beam direction.
It is covered by a movable concrete plate of 2.25~m thickness. Thus, 
depending on the point of impact on CMS, the total material traversed by
close-to-vertical muons changes from approximately 6 to 175 meters of water equivalent. 

The CMS experiment uses a right-handed coordinate system, with the origin at the
nominal proton-proton collision point, the $x$ axis pointing towards the center of the
LHC ring, the $y$ axis pointing upwards (perpendicular to the LHC plane), and
the $z$ axis pointing along the anticlockwise beam-direction,
at geographic azimuth $280.8^\circ$ (approximately west).
The angle between the CMS $y$ axis and the zenith direction
is 0.8$^\circ$. This small difference is neglected in the analysis,
and the angle of the muons relative to the $y$ axis is used to represent
the zenith angle $\theta_\mathrm{z}$.

At the center of the detector, the magnetic field is parallel to the
central axis of the solenoid, which is aligned with the $z$ axis. Muon momenta are
reconstructed by measuring the curvature of the muon trajectory projected
on the $xy$ plane, which yields the component
of muon momentum transverse to the $z$ axis, $\PT = p \sin\theta$,
where $\theta$ is the polar angle with respect to the $z$ axis.
This configuration is favourable for the reconstruction
of atmospheric muons, providing a strong magnetic bending for muons
traversing the detector, at any
incident azimuthal angle $\phi$ around the $z$ axis.
Full tracking of muons is available in the polar angle range
$10^\circ <  \theta < 170^\circ$.

CMS collected cosmic ray data in several runs during the final years of
detector construction and commissioning. Data from
the Magnet Test and Cosmic Challenge in 2006 (MTCC)~\cite{mtcc} and the
Cosmic Run At Four Tesla in 2008 (CRAFT08)~\cite{CRAFT08}
are used in the analysis reported here.

In August 2006 the CMS detector was pre-assembled on the surface 
before being lowered into the cavern. In this configuration no material 
above the detector was present, apart from the thin metal roof of the
assembly hall. A small fraction of each of the subdetectors
was instrumented and operating at the time. The details of the MTCC
setup are described in~\cite{mtcc,CMS_NOTE_2008-016}.
About 25 million cosmic-muon events were recorded during the first
phase of the MTCC with the magnet at a number of field values ranging from 
3.67 to 4.00~T.

The CRAFT08 campaign was a sustained data-taking exercise in October and November 2008
with the CMS detector fully assembled in its final underground position.
The full detector, ready for collecting data from LHC,
participated in the run, with the magnet at the nominal field of 3.8~T. 
Approximately 270 million cosmic-muon events were recorded.

Single cosmic muons are simulated using the
Monte Carlo event generator CMSCGEN~\cite{cmscgen,CMScrNote331},
which makes use of parameterizations of the distributions of the muon energy
and incidence angle based on the air shower program CORSIKA~\cite{corsika}.
The CMS detector response is simulated using the GEANT4 program~\cite{geant},
which takes into account the effects of energy loss, multiple scattering,
and showering in the detector.
A map~\cite{CMScrNote331} describing the various materials
between the Earth's surface and the CMS detector
is used to obtain the average expected energy loss of
simulated muons as a function of their energy, impact point, and incidence direction
at the surface.

\section{Cosmic-muon reconstruction}
\label{cosmic-reconstruction}

Muon tracking in CMS can be performed with the all-silicon tracker
at the heart of the detector, and with either three or four stations
of muon chambers installed outside the solenoid,
sandwiched between steel layers serving both as hadron absorbers and as a 
return yoke for the magnetic field.

\begin{figure*}[ht]
   \centering
   \includegraphics[width=0.90\textwidth]{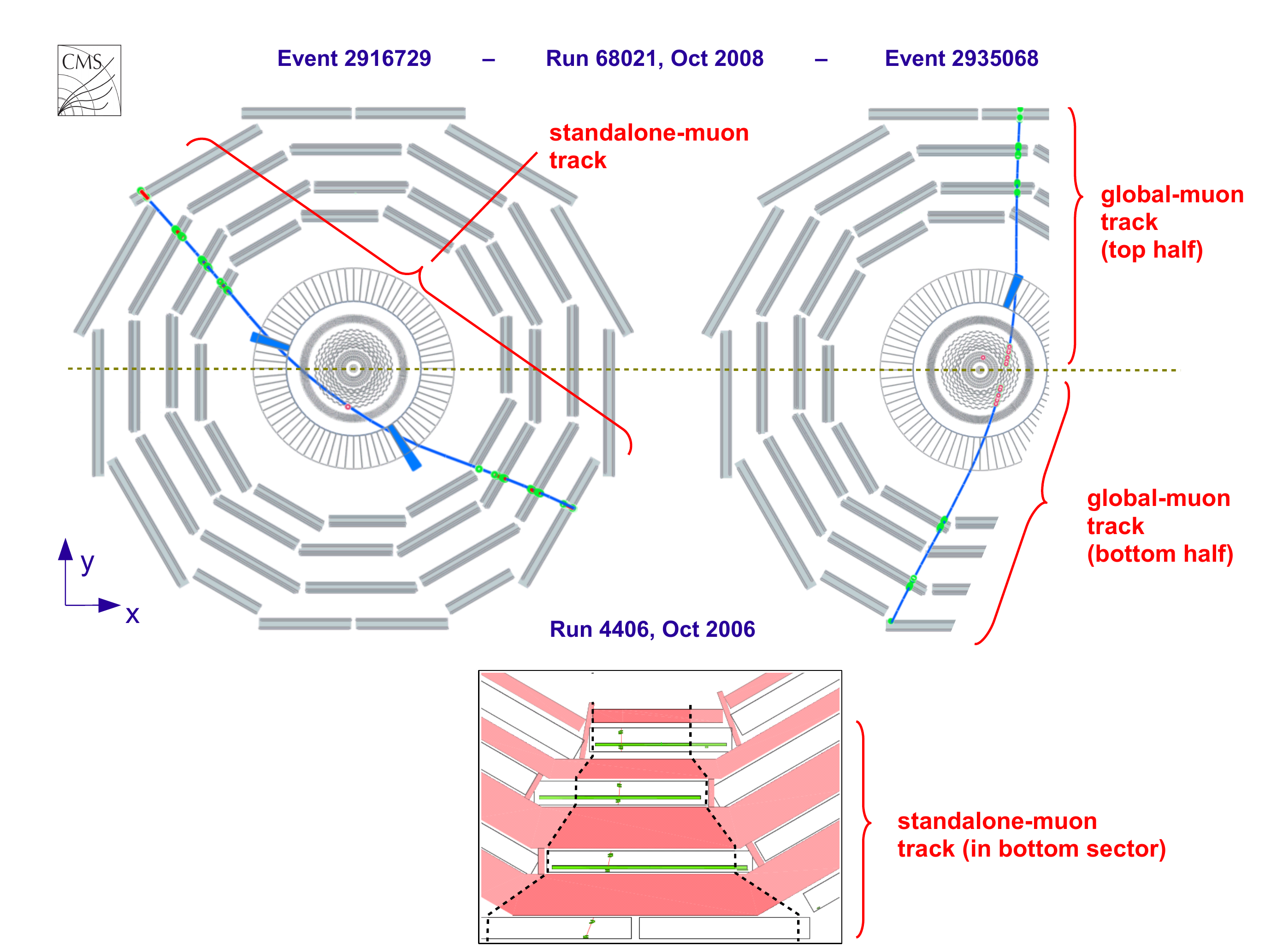}
   \caption{Cosmic-ray muons crossing the CMS detector.
            The upper two pictures display muons from 2008 underground data,
            leaving signals in the muon system, tracking detectors and calorimeters.
            A standalone track (top left) and a pair of global half-tracks (top right)
            are shown. The bottom plot depicts a muon from 2006 surface data crossing
            the muon chambers at the bottom of CMS.
   \label{fig:event-display}}
\end{figure*}

Three types of muon-track reconstruction were designed
for cosmic muons not originating from
an LHC proton-proton collision~\cite{CFT-09-014}:
a standalone-muon track includes only hits from the muon detectors;
a tracker track includes only hits from the silicon tracker;
and a global-muon track combines hits from the muon system and the 
silicon tracker in a combined track fit.
For a cosmic muon that crosses the whole CMS detector, illustrated in
Fig.~\ref{fig:event-display}~(top), each of the above types of 
tracks can be fitted separately in the top and bottom halves of CMS. 
Alternatively, a single track fit can be made including hits
from the top and bottom halves of CMS.
The direction of the muon is assumed to be downwards, and the muon charge
is defined accordingly.

The analysis based on 2006 MTCC data uses standalone muons. 
The reduced detector setup used in the MTCC was just a fraction of the bottom half 
of the complete detector, depicted in Fig.~\ref{fig:event-display}~(bottom).
Since the muons were measured only in one
half of the detector, the momentum resolution is poorer than in the standalone-muon
analysis using the complete detector. Having the detector on the surface, however, 
permitted the collection of a large number of low-momentum muons, 
down to a momentum of 5\GeVc{}, 
allowing for a precise measurement of the charge ratio in the low-momentum range.

Two analyses based on the 2008 CRAFT08 underground data are performed,
one using standalone muons and the other using global muons.
The underground global-muon analysis (GLB) profits from the
excellent momentum resolution and charge determination of global-muon tracks, 
but requires that the muon passes through the silicon tracker.
The underground standalone-muon analysis (STA) profits from the 
larger acceptance of the muon chambers and
yields approximately eight times as many muons as the global-muon analysis.
In a standalone cosmic-muon fit spanning the
whole diameter of the muon detector (Fig.~\ref{fig:event-display}), the momentum resolution
is significantly improved compared to a standalone fit using
only half the detector.
The improvement varies from a factor of four at low
momentum to more than a factor of ten for momenta above 100~\GeVc{}~\cite{CFT-09-014}.

\begin{figure*}[ht]
   \centering
   \subfigure[]{\includegraphics[width=0.49\textwidth]{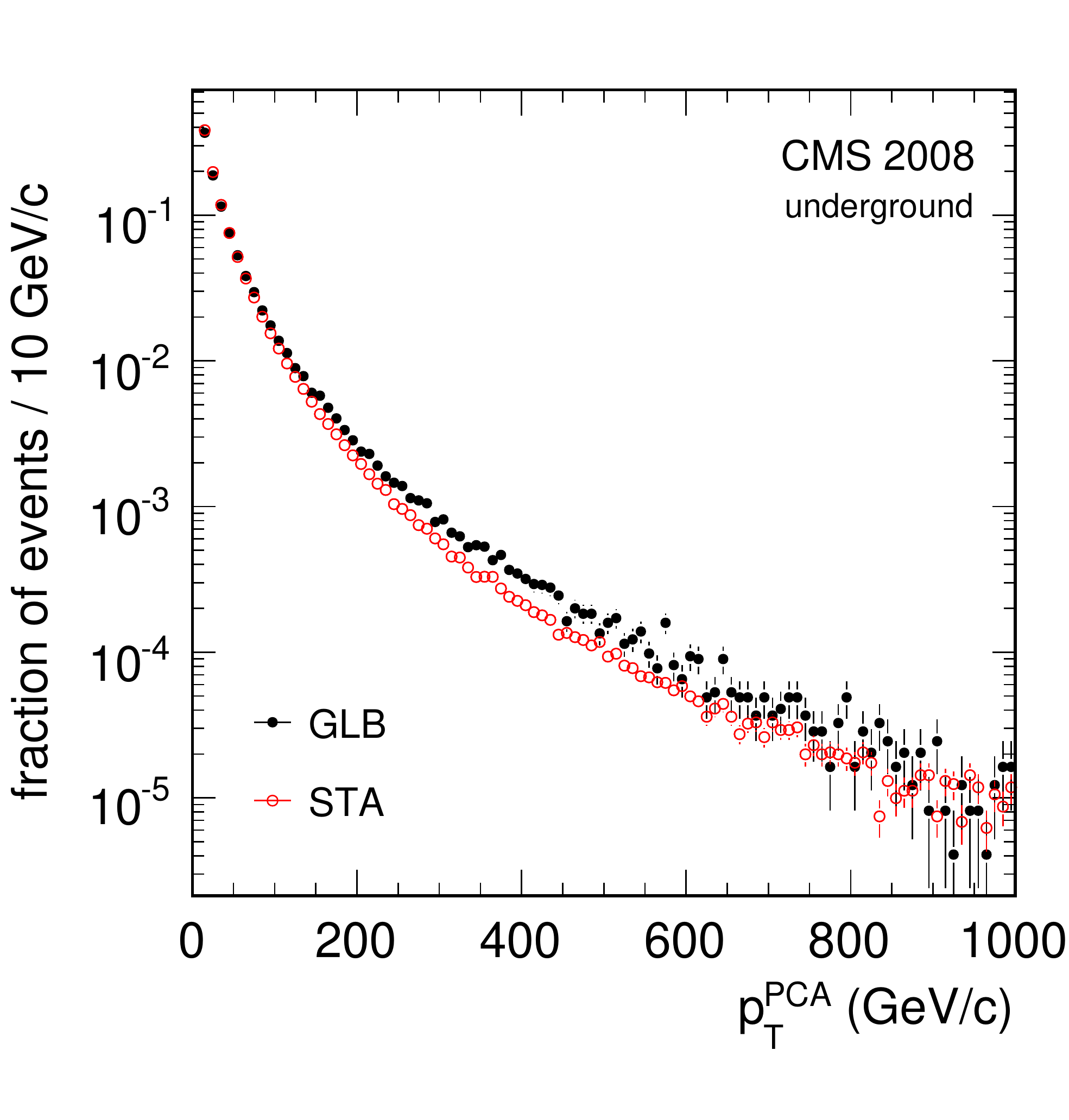}}
   \subfigure[]{\includegraphics[width=0.49\textwidth]{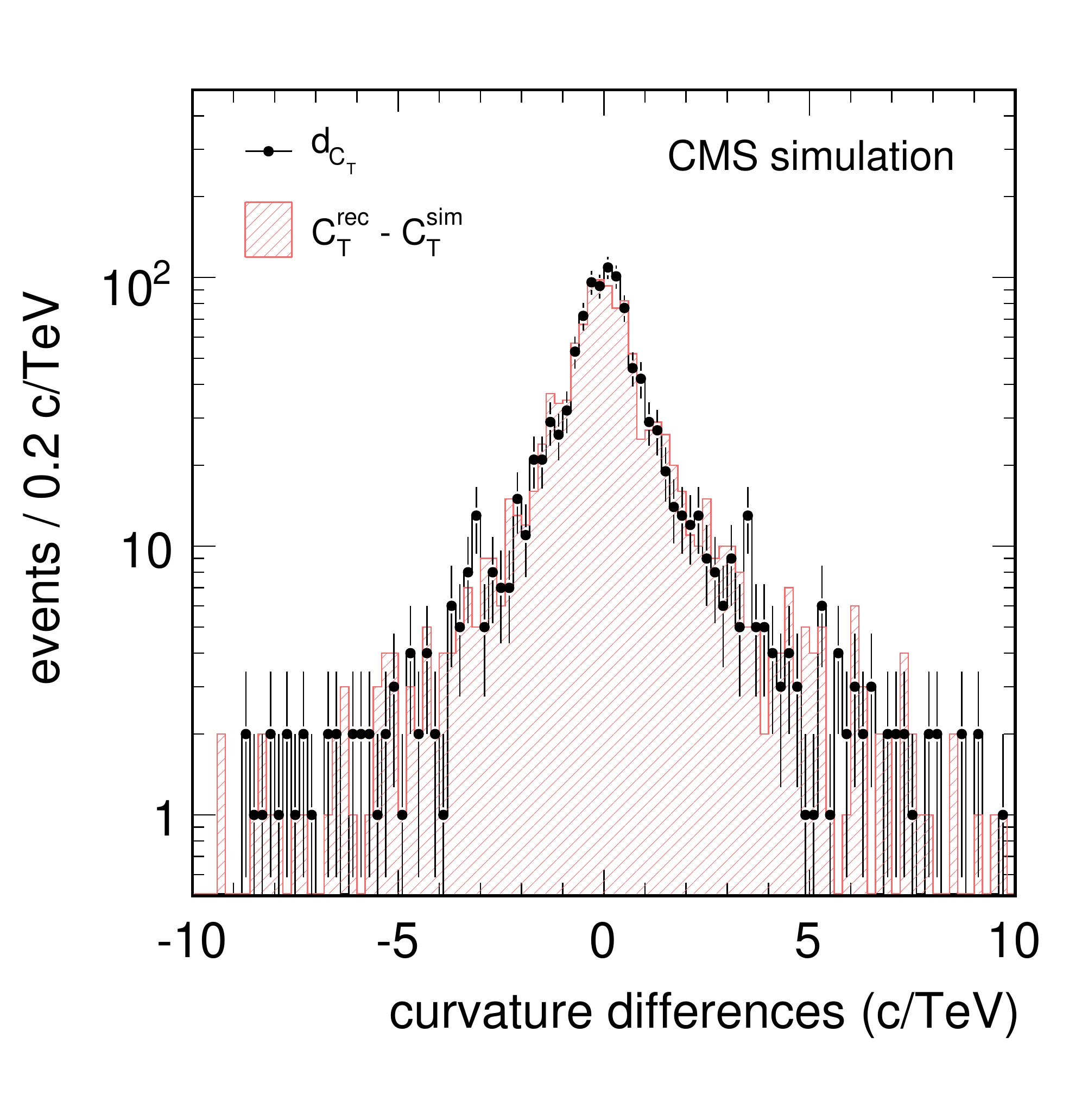}}
   \caption{(a) Normalized muon $\PT$ distributions, for the global (closed circles)
            and standalone-muon analyses (open circles), at the PCA.
            Differences in the distributions are expected, as the global and
            standalone-track fits have different momentum resolutions and acceptances.
            (b) Comparison of the ($q/\PT$) resolution estimate $d_{C_\mathrm{T}}$
            (closed circles) with the true $C_\mathrm{T}$ resolution (hatched histogram),
            obtained from simulated global muons.}
   \label{fig:GLB-splitMuResolution}
\end{figure*}

The ``maximum detectable momentum'', $p_\mathrm{mdm}$, defined as the momentum for which
the curvature of a muon track is measured to be one standard deviation away
from zero, is around $200 \GeVc{}$ for standalone-muon tracks
in one half of the detector,
around $10 \TeVc{}$ for standalone-muon tracks traversing the entire detector,
and in excess of $20 \TeVc{}$ for global-muon tracks.
The distribution of the transverse momentum ($\PT^\mathrm{PCA}$),
calculated at the point of closest approach (PCA)
to the nominal proton-proton collision point
taking into account the energy loss in the detector,
is depicted in Fig.~\ref{fig:GLB-splitMuResolution}~(a) for the
muons selected in the global and standalone-muon underground analyses.

The redundancy of the different tracking systems in the complete CMS detector
allows the determination of the momentum resolution and rate of charge misassignment
(the fraction of muons reconstructed with incorrect charge)
directly in data. In the global-muon analysis, the 
half-difference of the track curvatures measured in the top half
and the bottom half of the detector $d_{C_\mathrm{T}}$ is used to measure
the resolution of the half-sum $C_\mathrm{T}$:
\begin{eqnarray}
\label{eq:resol}
C_\mathrm{T} & \equiv & \left(\frac{q}{\PT}\right)_\mathrm{average} = \frac{1}{2}\left[  \left( 
\frac{q}{\PT} \right)_\mathrm{top} + \left( \frac{q}{\PT}\right)_\mathrm{bottom} \right ], \quad
\nonumber \\
d_{C_\mathrm{T}} & \equiv & \frac{1}{2}\left[  \left( 
\frac{q}{\PT} \right)_\mathrm{top} - \left( \frac{q}{\PT}\right)_\mathrm{bottom} \right ],
\end{eqnarray}
where \PT{} is the transverse momentum and $q$ the charge sign of the muon.
Both the core and the tails of the resolution distribution are
well reproduced by the $d_{C_\mathrm{T}}$ estimator, as demonstrated for simulated events
in Fig.~\ref{fig:GLB-splitMuResolution}~(b).

In the underground standalone-muon analysis, an independently reconstructed tracker track
is available in 40\% of the selected events. The comparison of
the charge and momentum measured for the tracker track and for the standalone-muon track
gives a measure of the tracking resolution both in data and in simulated events.

All three analyses measure the charge ratio in events with a single cosmic ray muon,
rejecting events with more than one muon detected.

\section{Event selection and analysis}
\label{event-selection}

\subsection{Analysis of surface data}

The cosmic-muon charge ratio was measured by CMS for the first time using
MTCC data~\cite{CMS_NOTE_2008-016}.
For this analysis, only the bottom sector in two (out of five) wheels of the barrel
muon system (DT) is used. Selection accepts
only muons triggered and reconstructed in a perfectly 
left-right symmetric fiducial volume with respect to the vertical axis,
emphasized in Fig.~\ref{fig:event-display}~(bottom),
ensuring a charge-symmetric acceptance. 

\begin{figure*}[ht]
   \centering
   \subfigure[]{\includegraphics[width=0.49\textwidth]{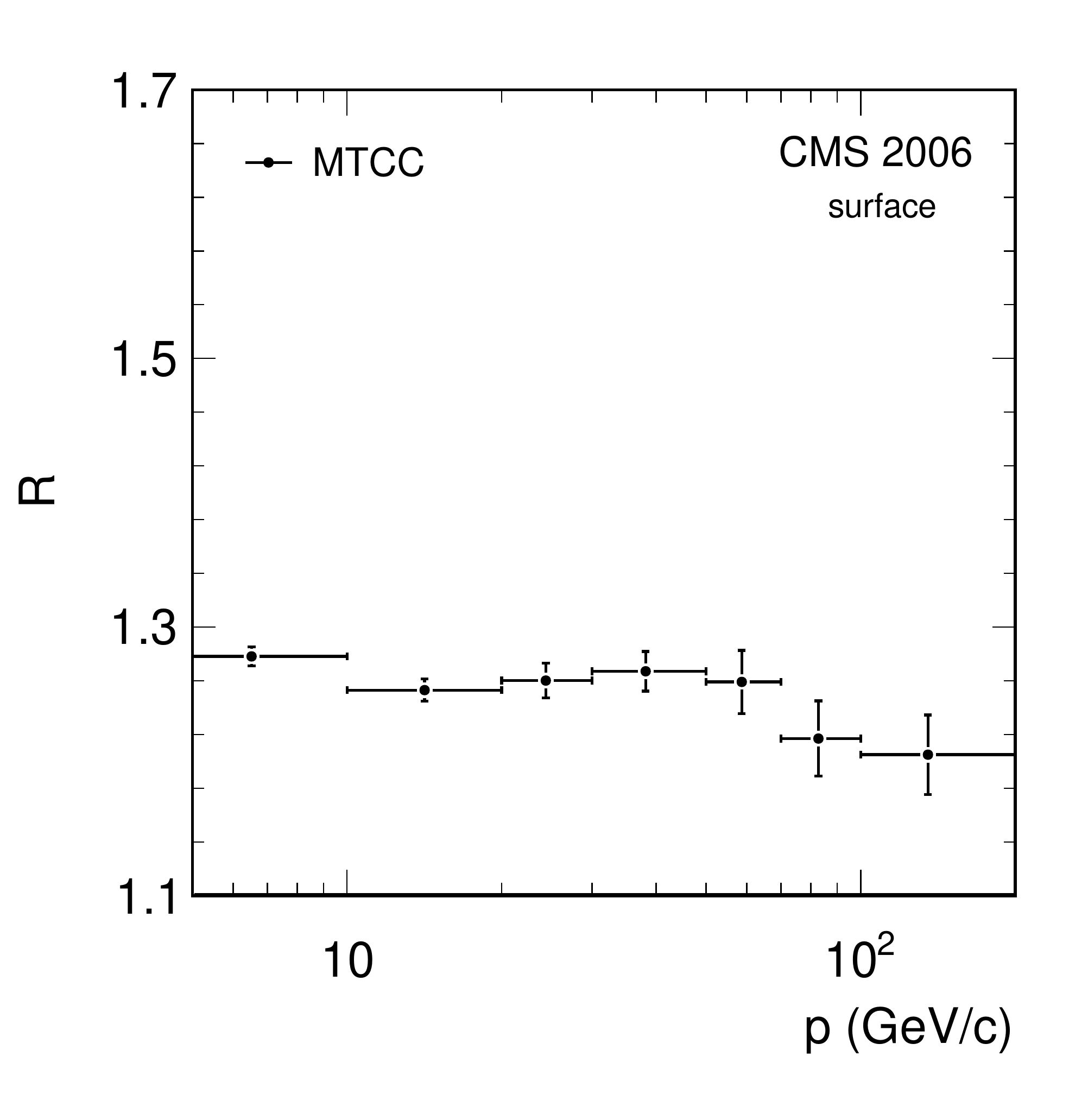}}
   \subfigure[]{\includegraphics[width=0.49\textwidth]{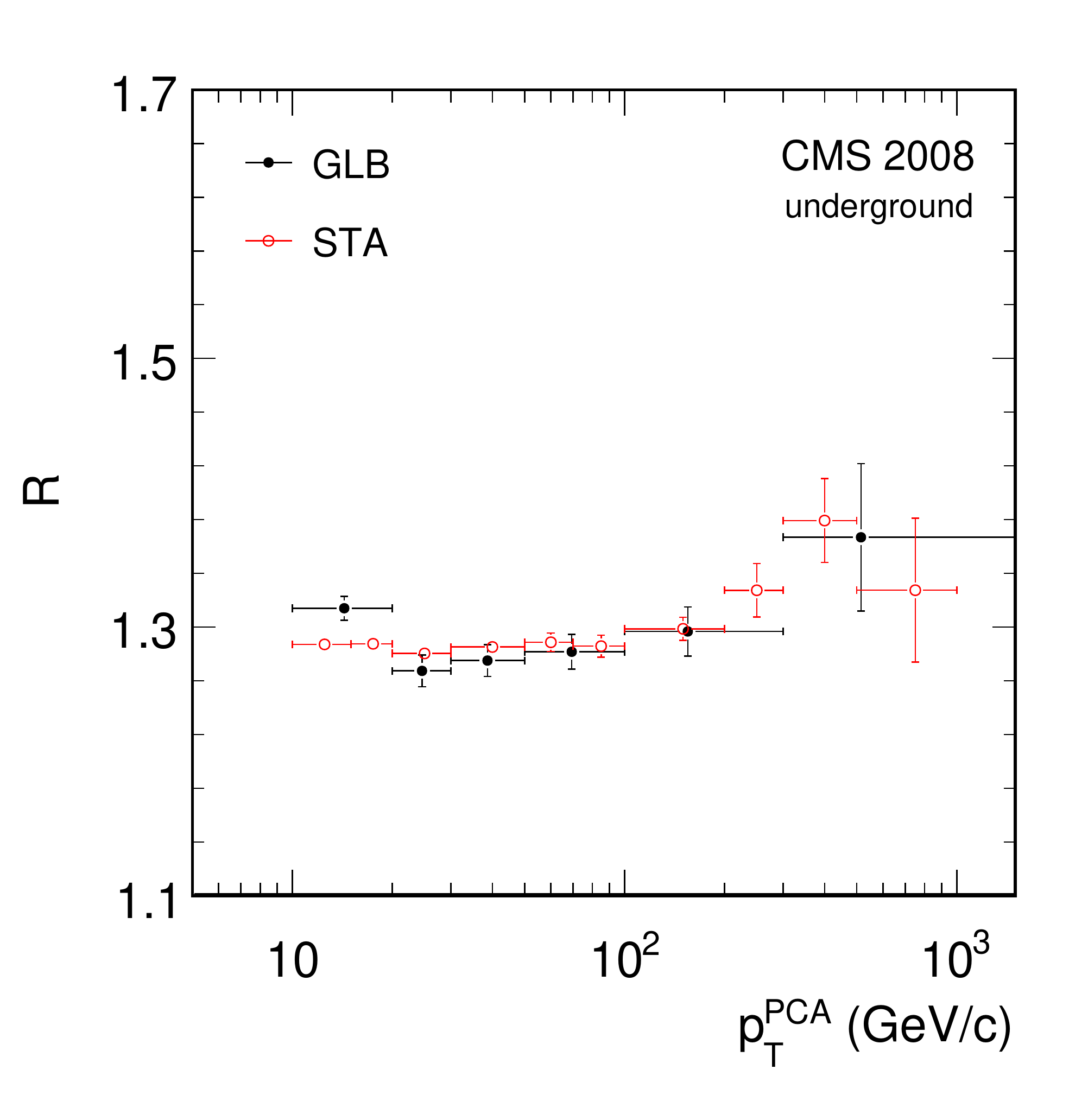}}
   \caption{Uncorrected charge ratio, together with the statistical uncertainty.
            (a) From 2006 MTCC data, as a function of the measured muon momentum.
            (b) For the global (closed circles) and standalone-muon
            analyses (open circles),
            as a function of the measured $\PT$ at the~PCA.
   \label{fig:raw-charge-ratio}}
\end{figure*}

Around 15 million events were recorded in runs with 
DT triggers and a stable magnetic field above 3.67~T. About
$330\,000$ events pass the fiducial-volume
and track-quality selections.
The measured muon charge ratio and its statistical uncertainty
are displayed in Fig.~\ref{fig:raw-charge-ratio}~(a), as a function
of the measured muon momentum,
before any correction due to detector effects is applied.

The probability of charge misassignment is small for low-momentum muons.
At high momenta, resolution effects increase the chance of charge misassignment
thus lowering the measured value of the charge ratio.
Only muons with a measured momentum below $p_\mathrm{mdm} =
200 \GeVc{}$ are included in the analysis.

\subsection{Underground global-muon analysis}

The 2008 data were recorded using a single-muon trigger requiring the
coincidence of muon hits in at least two muon detector layers. Triggers from
the DT or RPC systems in the top or the bottom halves of the detector
were accepted. The trigger efficiency is high for muons with sufficient
momentum (a few~\GeVc{}) to penetrate several layers of the steel return
yoke~\cite{L1triggerCraft}.
The subsequent event selection is designed to ensure good track quality and high
efficiency.

The muon trajectory in each half of the detector is required to contain 
at least 20 (out of 44 possible) hits in the DT system. Of these 20 hits, at least 3 hits
are required to measure the longitudinal coordinate ($z$), ensuring a good
measurement of the polar angle. The muon trajectory is required to contain 
no hits in the muon or tracker endcaps. The two halves, top and bottom, of each cosmic-muon
trajectory are required to be reconstructed as two
separate track segments in the silicon tracker, each containing
at least 5 hits (out of 12 possible) in the tracker outer barrel system.
A loose cut is applied to the normalized $\chi^2$ of each of the two global-muon fits and the
polar angles are required to match within $| \Delta \cot \theta | < 0.2$,
in order to suppress the small background from multi-muon cosmic shower events.
The average transverse momentum of each muon, measured at the
PCA, is required to be greater than 10\GeVc{}
in order to ensure that the muon is able to traverse the entire CMS detector.
All selection requirements are applied to the top and bottom muon trajectories.

While the main shaft of the CMS underground area is symmetric
with respect to the $yz$ plane, the two auxiliary access shafts are 
located at asymmetric positions with respect to this plane
(cf.~Section~\ref{sec:experiment}).  This causes
the geometrical acceptance of the detector to be asymmetric
for muons of different charges, since the CMS magnetic field 
is aligned with the $z$ axis. To remove this effect, muon tracks that
cross these auxiliary shafts are not considered in the analysis, 
nor are muons that cross the mirror images of those regions
with respect to the $x=0$ plane. 
We refer to this requirement as ``symmetric selection''. 

About $245\,000$ muons are selected. The muon $\PT$
distribution is reported in Fig.~\ref{fig:GLB-splitMuResolution}~(a) for the
selected muons. Figure~\ref{fig:raw-charge-ratio}~(b) depicts
the measured uncorrected charge ratio as a
function of~$\PT^\mathrm{PCA}$.

\subsection{Underground standalone-muon analysis}
\label{section:STA_analysis}

In this analysis the particle trajectory is reconstructed using
only the hits in the barrel muon system (DT and RPC).
To select muon tracks that are fully contained in the
barrel region, events with hits in the endcap CSCs are rejected.
A single track is reconstructed using the information from both halves of the detector.
Only one standalone muon per event is allowed.

Muon tracks are required to have a transverse momentum, measured at the PCA,
larger than 10~\GeVc{}.
At least 45 muon hits (out of 88 possible) are required to be associated with the track.
The muon trajectory in the event is also reconstructed as two
standalone-muon tracks, one in the upper and one in the lower half of
the detector, with more than 20 hits (out of 44 possible) each.

In order to ensure a good track-quality, further
selection criteria are applied to the tracks:
the normalized $\chi^2$ of each reconstructed muon track must be less 
than 5,
the impact parameter in the $xy$ plane must be less than 100~cm,
the track direction at PCA must be vertical within $42^\circ$ in $\theta$
and $60^\circ$ in $\phi$,
and the track PCA must lie within the range $|z| < 600$~cm.
A ``symmetric selection'' is also applied as in the global-muon analysis.
The number of muons selected is 1.6~million.

The analysis relies on the simulation to correct for charge misassignment
and momentum resolution effects, using the data with both a standalone and a tracker
track in the event to perform further corrections and estimate systematic uncertainties.
From the comparison of tracks reconstructed both in the tracker and in the muon
system, the probability of charge misassignment is known to be well below 1\% for
$\PT^\mathrm{PCA} < 0.5 \TeVc$,  increasing up to about 1.5\%
in the highest momentum bin.
The difference observed between data and simulation in the subsample of
events that include a tracker track is taken into account to correct the
charge misassignment and to assign the related systematic uncertainty,
as explained in Section~\ref{section:systematics}.

The muon momentum scale and resolution are determined by comparing the
transverse momentum of the standalone-muon track to that of the
associated tracker track, and are accurately modeled
by the simulation. Therefore the momentum unfolding, which provides
an estimate of the true momentum of the muon tracks from the measured
momentum, can be based on the simulation. An uncertainty on
the momentum resolution for all events, including those without a tracker
track, is taken into account as a systematic uncertainty. The momentum scale
in the tracker volume is set by the magnetic field, which is known to a
precision better than 0.1\%~\cite{BfieldCRAFT}, as confirmed by additional
checks performed with early LHC data~\cite{TRK-10-001}. 
The uncorrected muon charge ratio is shown
in Fig.~\ref{fig:raw-charge-ratio}~(b) as a function of~$\PT^\mathrm{PCA}$.

\section{Corrections for energy loss and resolution}
\label{Extrapolation-and-unfolding}

In order to express the charge ratio measurement as a function of the 
true momentum at the surface of the Earth,
the measured momentum inside the CMS detector has to be corrected for energy lost
between the surface of the Earth and the point of measurement. Furthermore, corrections 
need to be applied for migration of entries from bin to bin due to momentum resolution 
and for possible misassignment of the muon charge.

\subsection{Energy-loss correction}

In the MTCC analysis the measured muons are propagated back to the top of CMS, correcting
for expected momentum loss and bending in the magnetic field. In addition, the effect of 
charge misassignment is estimated using simulated events, and a bin-by-bin correction 
is applied to the measured charge ratio.

For the muons selected in the global and standalone-muon analyses of the 2008 underground
data, the average expected energy loss 
depends strongly on the path followed through the Earth.
The underground measurements are corrected for this effect by 
propagating the trajectory of individual muons back to the 
Earth's surface, using
the same material model as in the simulation (cf.\ Section~\ref{sec:experiment}).
Energy loss in matter is about 0.15\% higher for $\mu^+$
than for $\mu^-$ due to slightly larger ionization losses~\cite{arXiv09063726}.
This difference is taken into account in the energy-loss correction,
but affects the measured charge ratio by less than 0.3\% over the entire momentum range.

\subsection{Unfolding the momentum spectrum\label{unfolding-muon-spectrum}}
\label{section:unfolding}

In the underground data analyses, momentum resolution effects in the detector
are corrected using an unfolding technique, applied to the charge-signed inverse momentum $C = q/p$.
In this procedure $p$ represents the measured momentum extrapolated to the Earth's 
surface, where the correlation
with the true muon momentum is highest.

The momentum measured at the PCA is propagated
first to the top of CMS, accounting for the magnetic field and the amount
of material traversed, and then from the top of CMS to the
surface of the Earth, following a straight line. The angular resolution of the
detector is better than 5~mrad. Only muons with an 
estimated momentum above 30\GeVc{} after this correction are kept in the analyses.

Given a vector of true muon counts $N_{j}^\mathrm{true}$
matrix inversion is used to compute the best estimator
$\widetilde{N}_i^\mathrm{true}$ from the vector of observed muon counts ${N}^\mathrm{measured}_i$:
\begin{eqnarray}
\label{eq:one-shot-unfolding}
N_{i}^\mathrm{measured} & = & \sum_j M_{ij} N_{j}^\mathrm{true} \, , \quad \nonumber \\
\widetilde{N}_i^\mathrm{true} & = & \sum_j \widetilde{M}_{ij}^{-1} N_{j}^\mathrm{measured} \, .
\end{eqnarray}
The migration matrix element $M_{ij}$ is the probability
that a muon with true $C$ ($C^\mathrm{true}$) in bin $j$ is observed with
a measured $C$ ($C^\mathrm{measured}$) in bin $i$. $\widetilde{M}_{ij}$
is an approximation of the exact migration matrix, and is
constructed differently for the global and standalone-muon analyses.

In the standalone-muon analysis the migration matrix estimator is extracted by 
comparing the true momentum to the reconstructed momentum in simulated events.

In the global-muon analysis the approximate migration matrix is 
derived directly from the data. For each muon, the $C$ values measured 
in the top and the bottom half of the detector are propagated individually 
to the Earth's surface. 
The estimated true $C$ is then defined as $\widetilde{C}^\mathrm{true} = ( C_\mathrm{top} + C_\mathrm{bottom}) / 2$,
and the measured values $C_\mathrm{top}$ and $C_\mathrm{bottom}$ are used to represent $\widetilde{C}^\mathrm{measured}$. 
They both have the desired property
$\widetilde{C}^\mathrm{measured} = \widetilde{C}^\mathrm{true} \pm d_C$, where $d_C$ is the $C$ resolution estimator,
defined as $d_C = (C_\mathrm{top} - C_\mathrm{bottom}) / 2$.
The matrix $\widetilde{M}_{ij}$ is then populated using these estimated values,
for all muons in the selected event sample. As the resolution
estimator $d_C$ gives a good representation of the actual
resolution of $C^\mathrm{true}$ (Fig.~\ref{fig:GLB-splitMuResolution}~(b)), this
procedure yields a good approximation of the true migration matrix $M_{ij}$.

In both analyses, variations of the energy loss around the expected value are
taken into account in the unfolding procedure by applying an additional 10\%
Gaussian smearing of the energy-loss correction
to the measured momentum when forming the migration matrix.
This approximation is based on simulation studies using GEANT4.

The muon counts $N_{i}$ correspond to the bins of the histograms in which
the corrected charge ratio results are presented. The bin boundaries 
were chosen such that the migration between bins is small. The values of 
the off-diagonal elements of the migration matrix are below 0.1 in the
global-muon analysis and less than 0.2 in the standalone-muon analysis.

\begin{figure}[ht]
   \centering
   \includegraphics[width=0.49\textwidth]{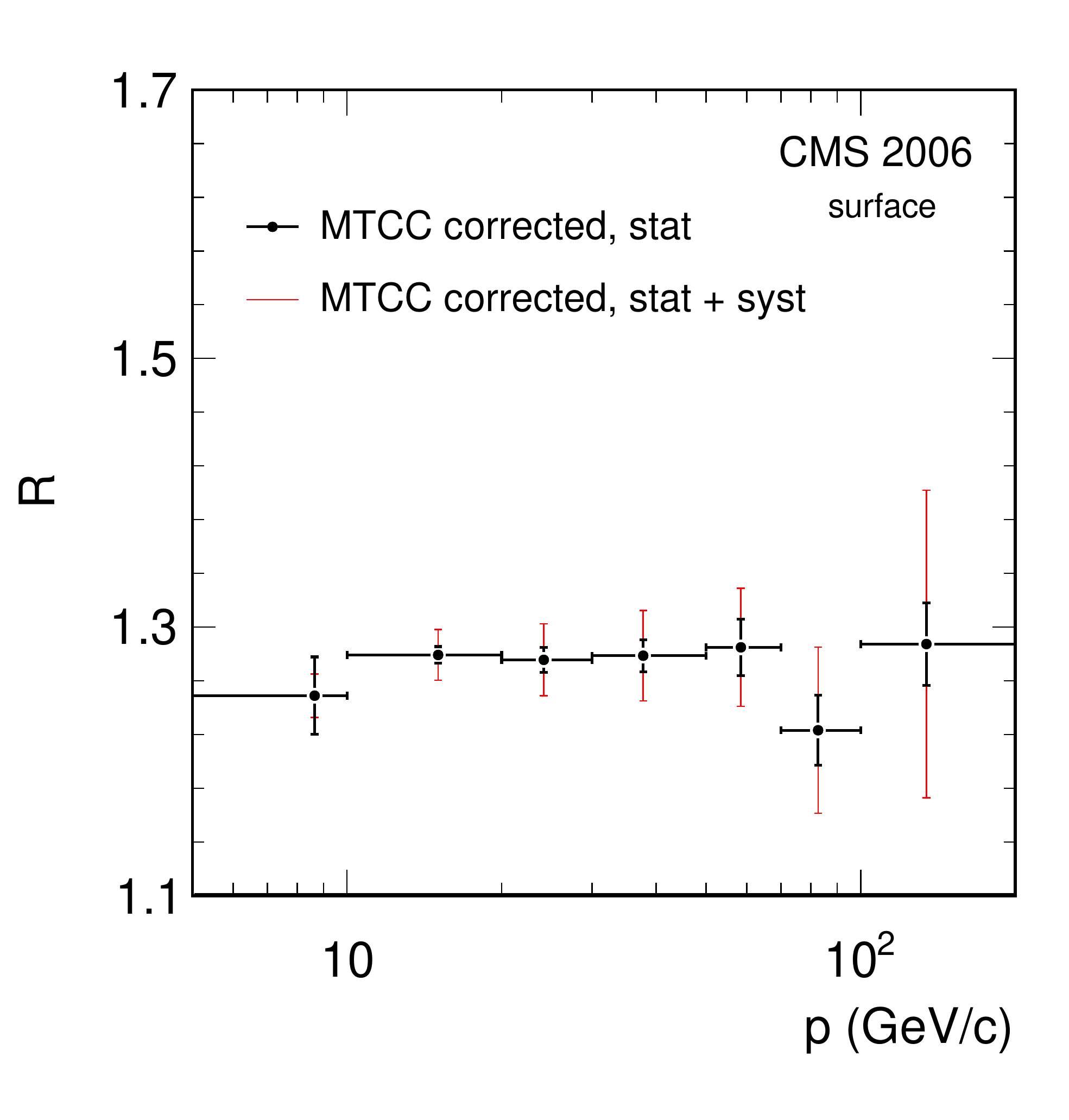}
   \caption{Charge ratio for the surface analysis, as a function of the muon momentum,
            corrected for energy loss in the detector and for charge misassignment,
            after propagating the muon track to the entry point in CMS.
            The thick error bars denote the statistical uncertainty and
            the thin error bars statistical and systematic uncertainties
            added in quadrature.
   \label{fig:MTCC-Rcorr}}
\end{figure}

The measurement of the charge ratio using
2006 data, corrected for energy loss in the detector and for charge misassignment,
is depicted in Fig.~\ref{fig:MTCC-Rcorr} as a function of the muon momentum,
together with the statistical and systematic uncertainties.

\begin{figure*}[ht]
   \centering
   \subfigure[]{\includegraphics[width=0.49\textwidth]{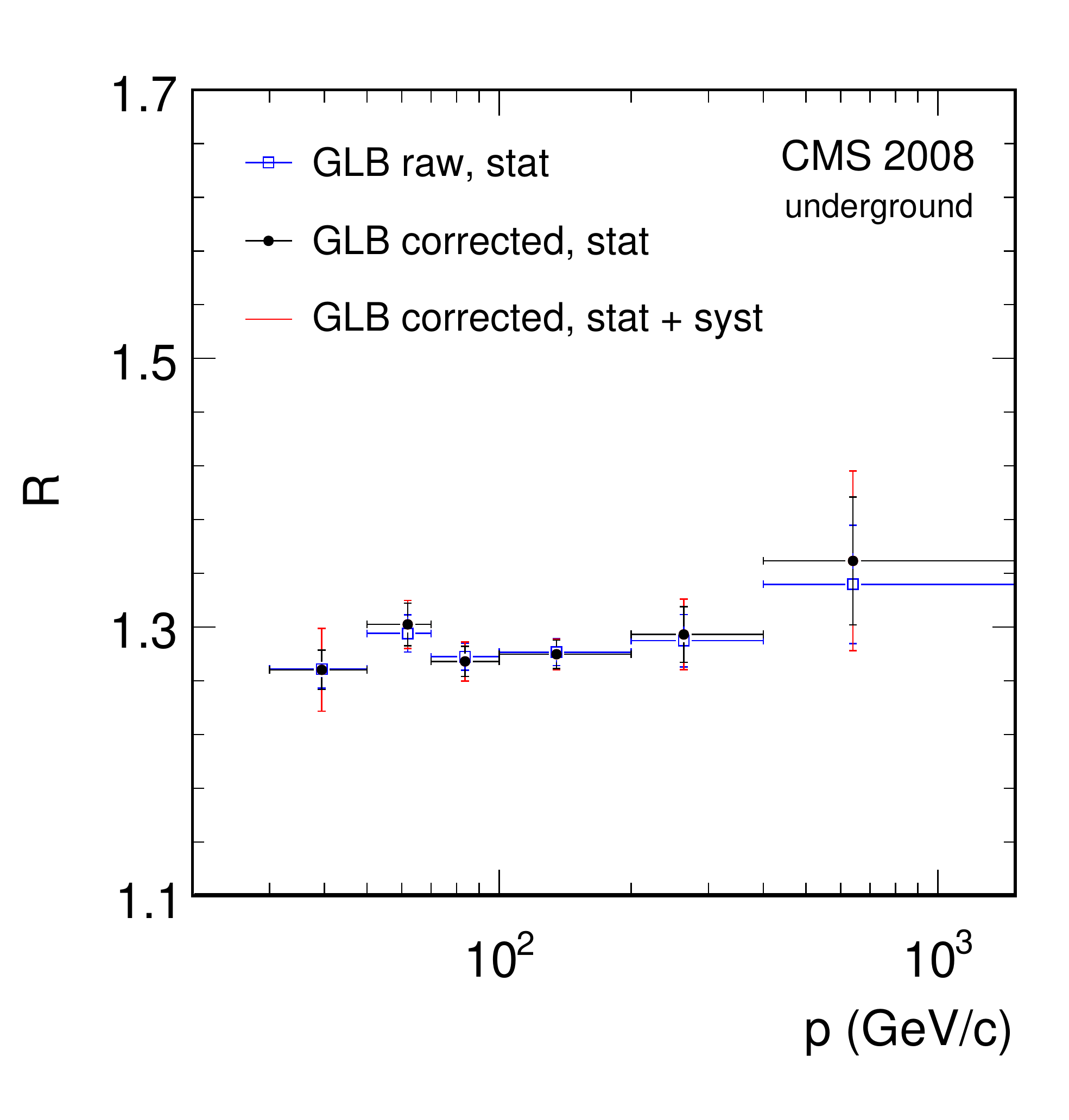}}
   \subfigure[]{\includegraphics[width=0.49\textwidth]{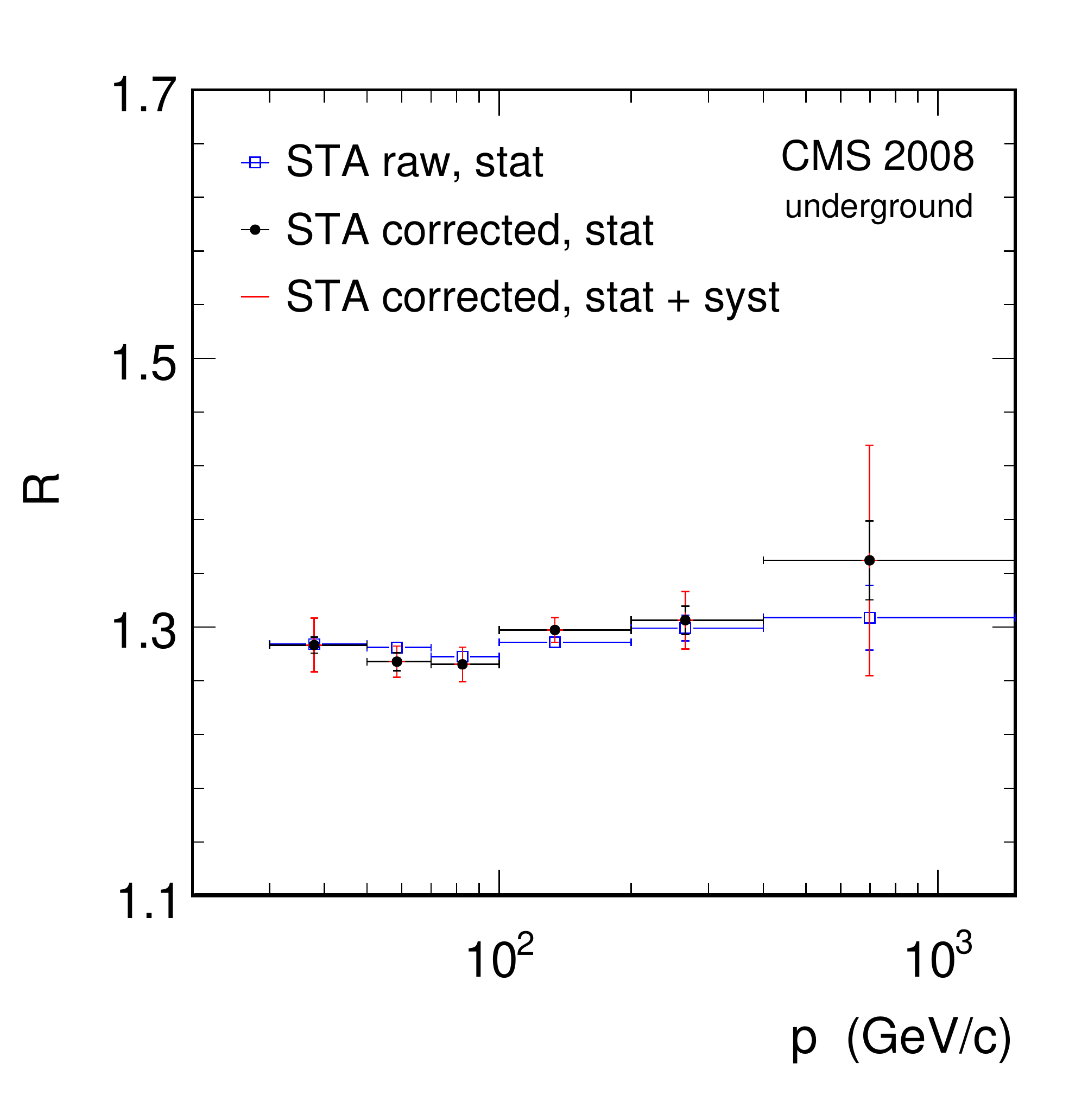}}
   \caption{Muon charge ratio as a function of the muon momentum at the
            Earth's surface for (a) global and (b) standalone-muon analyses.
            Open squares indicate the uncorrected ratio, including full alignment.
            Closed circles show the unfolded charge ratio with statistical
            errors only. The lines denote the statistical and systematic
            uncertainties added in quadrature.
   \label{fig:RES-CMSunfolded}}
\end{figure*}

The measurements of the muon charge ratio in the global and 
standalone-muon analyses of 2008 data are displayed in Fig.~\ref{fig:RES-CMSunfolded},
as a function of the muon momentum.
The ``raw'' result is based on the final alignment including the
scale correction discussed in Section~\ref{section:systematics}. The ``corrected''
results are based on the unfolding and, for the standalone-muon analysis, include
an additional charge-misassignment correction.

\section{Systematic uncertainties}
\label{section:systematics}

Systematic uncertainties arise from reconstruction and instrumental effects that
can affect differently the detection efficiency and momentum measurement
of $\mu^+$ and $\mu^-$. They are evaluated as a function of the muon momentum
at the Earth's surface.

The CMS magnetic field is known with high precision in the region inside
the superconducting solenoid, and with less precision in the steel return
yoke~\cite{BfieldCRAFT}.
Systematic effects on the charge ratio due to the uncertainty on the
magnetic field are less than 1\%.

A possible bias in the positive and negative muon rates detected underground, due to
asymmetries in detector acceptance and uncertainties in the material densities used in
the material map (known within 5\%),
yields a non-negligible uncertainty on the charge ratio only in the lowest momentum bin.
The additional effect of the selection cuts is generally small, well below~1\%.

The requirement of a muon trigger in the detector leads to a small difference in
efficiency for positive and negative muons, below 1\%, which is correlated 
between the two underground analyses.
Both analyses estimate a possible systematic bias induced by the trigger
by employing a so-called tag-and-probe technique, using information from both
halves of the detector and, in the case of the standalone-muon analysis,
information from the independent DT or RPC muon triggers.

In the global-muon analysis the effect of charge misassignment is small,
ranging from less than 0.01\% at 10~\GeVc{} to about 1\% at 500~\GeVc{}, and it is
corrected by the unfolding procedure, using the 
data-driven resolution estimator defined in Eq.~(\ref{eq:resol}).

In the standalone-muon analysis
the charge misassignment correction to the charge ratio, included in the
unfolding matrix, is based on simulated events and
tested in real data using the subsample of standalone
muons with an associated tracker track. A higher rate of charge misassignment 
is observed in data than in simulation,
with a maximum absolute discrepancy of 3\% in the highest
momentum bin. Since this discrepancy could not be attributed unambiguously to the 
standalone-muon tracks, a correction is applied equal to 50\% of the full effect
observed in data, with a systematic uncertainty equal to the correction itself.

The precise alignment of all the tracking-detector components
is crucial for accurate reconstruction of high-$\PT$ muons,
whose trajectories have only a small curvature in the detector.
Cosmic muon tracks from the same 2008 data set used
for this analysis are employed to perform such an
alignment of the silicon tracker and muon
system~\cite{CFT-09-003,CFT-09-016}.
Possible effects from potential residual
misalignment that could lead to momentum migrations
and incorrect charge assignments are evaluated
by studying various realistic missalignment scenarios in data and simulation.
Only the two highest momentum bins are potentially affected
by misalignment, as expected, yielding a bias in the charge ratio
around 1\% in the two highest-momentum bins for the global-muon analysis.
For the standalone-muon analysis,
the effect in the charge ratio is less than 1\% up to 400\GeVc{}, and around 4\%
in the highest-momentum bin.

A global deformation of the detector could be missed during the alignment procedures
(a so-called "$\chi^2$-invariant" or "weak" mode~\cite{WeakMode}),
and potentially affect the charge ratio. The most problematic 
deformation would be a mode which caused a constant offset in
$q / \PT^\mathrm{PCA}$, different from zero, affecting the momentum
scale for cosmic muons of opposite charge in opposite directions.
A two-parameter fit of the simulated $q / \PT^\mathrm{PCA}$ distribution
to the data is performed using muons
in the range $\PT^\mathrm{PCA} > 200 \GeVc{}$, leaving the unknown charge
ratio and the $q / \PT^\mathrm{PCA}$ offset in the simulation to vary freely in the fit.
An offset of $0.043 \pm 0.022$ $c$/{Te\hspace{-.08em}V} is found.
The measured muon momenta are corrected for this offset and its uncertainty
is included as an additional systematic uncertainty on $R$,
fully correlated between the two underground measurements,
of the order of 1\% and 4\% respectively in the two highest momentum bins.

In the 2006 MTCC analysis,\label{Syst-MTCC}
systematic uncertainties arise mainly from the finite precision of the
detector alignment parameters, from the correction of the charge
misassignment probability and from the slightly larger uncertainty ($\sim$5\%)
in the scale of the magnetic field in the steel return yoke.

The total systematic uncertainties in the three
analyses are summarized in Table~\ref{tab:systematics},
as a function of $p$ at the Earth's surface.
The systematic uncertainties have also been evaluated as a function of
the vertical momentum component, $p\cos\theta_\mathrm{z}$, an observable
on which the charge ratio is expected to depend in a simple way~\cite{arXiv09063726}.

\begin{table*}[ht]
\caption{Charge ratio $R$ and relative statistical (stat.) and systematic
(syst.) uncertainties in bins of $p$ (in \GeVc{}), for surface data and both
analyses of underground data. The relative uncertainties are expressed in \%.}
\label{tab:systematics}
\begin{center}
\begin{tabular}{r || c  c  c | c  c  c | c  c c }
\hline
\multicolumn{1}{c||}{$p$} & \multicolumn{3}{c|}{2006 surface}
                          & \multicolumn{3}{c|}{2008 global-muon}
                          & \multicolumn{3}{c}{2008 standalone-muon} \\
\multicolumn{1}{c||}{range} &  $R$  &  stat. &  syst. &  $R$  &  stat. &  syst. &  $R$  &  stat. &  syst. \\ \hline
  5 -- \pho 10 & 1.249 &  2.3   &  1.3   &  $-$  &  $-$   &  $-$   &  $-$  &  $-$   &  $-$   \\
 10 -- \pho 20 & 1.279 &  0.5   &  1.5   &  $-$  &  $-$   &  $-$   &  $-$  &  $-$   &  $-$   \\
 20 -- \pho 30 & 1.276 &  0.7   &  2.1   &  $-$  &  $-$   &  $-$   &  $-$  &  $-$   &  $-$   \\
 30 -- \pho 50 & 1.279 &  0.9   &  2.6   & 1.268 &  1.2   &  2.1   & 1.287 &  0.5   &  1.5   \\
 50 -- \pho 70 & 1.285 &  1.6   &  3.4   & 1.302 &  1.2   &  0.6   & 1.274 &  0.5   &  0.8   \\
 70 --     100 & 1.223 &  2.1   &  5.1   & 1.274 &  0.9   &  0.7   & 1.272 &  0.4   &  0.9   \\
100 --     200 & 1.287 &  2.4   &  8.9   & 1.280 &  0.8   &  0.3   & 1.298 &  0.3   &  0.6   \\
200 --     400 &  $-$  &  $-$   &  $-$   & 1.295 &  1.6   &  1.3   & 1.305 &  0.8   &  1.4   \\
\multicolumn{1}{c||}{$>400$}
               &  $-$  &  $-$   &  $-$   & 1.349 &  3.5   &  3.5   & 1.350 &  2.2   &  6.0   \\
\hline
\end{tabular}
\end{center}
\end{table*}

\section{Results}

\begin{figure*}
   \centering
   \subfigure[]{\includegraphics[width=0.49\textwidth]{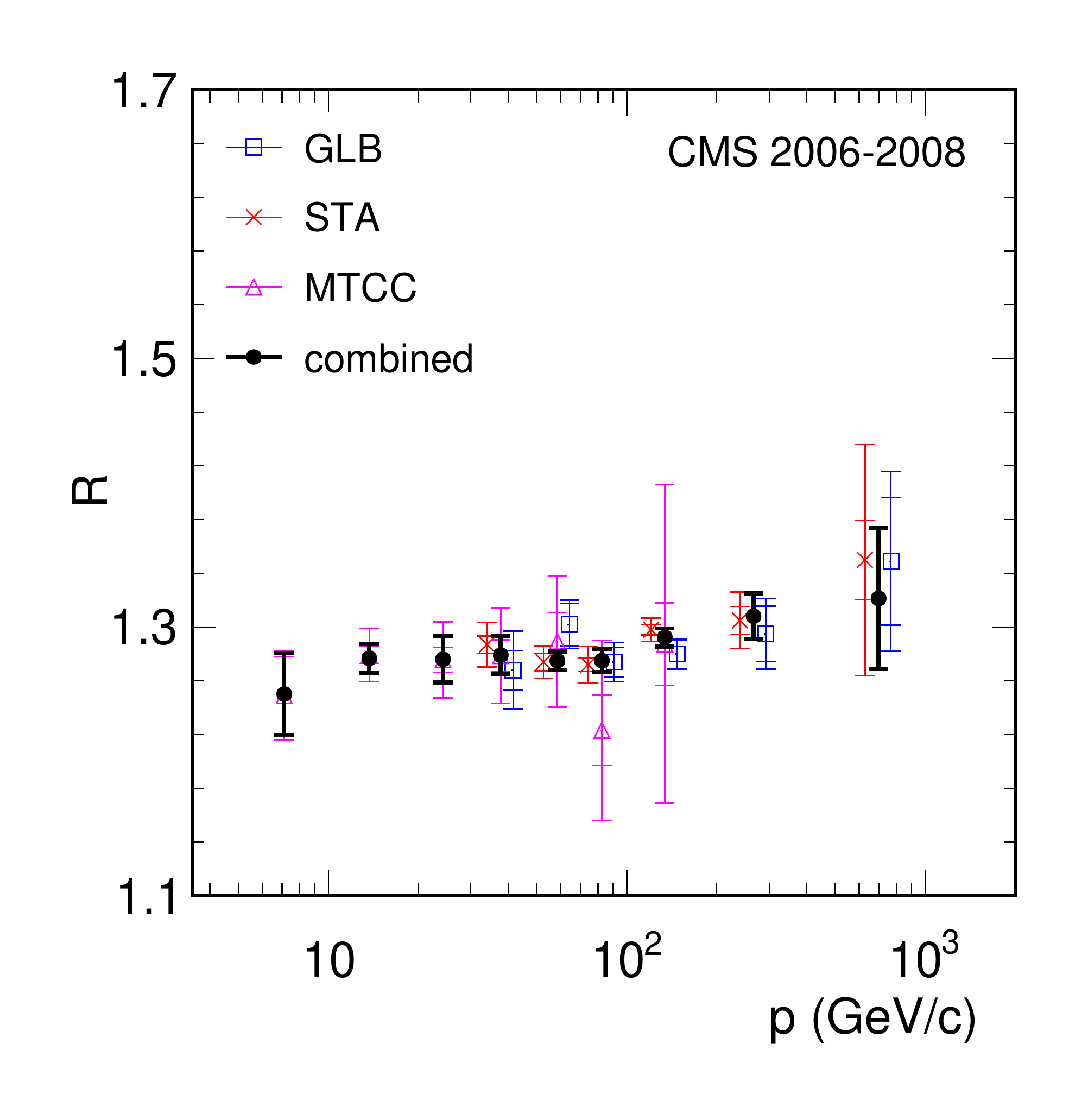}}
   \subfigure[]{\includegraphics[width=0.49\textwidth]{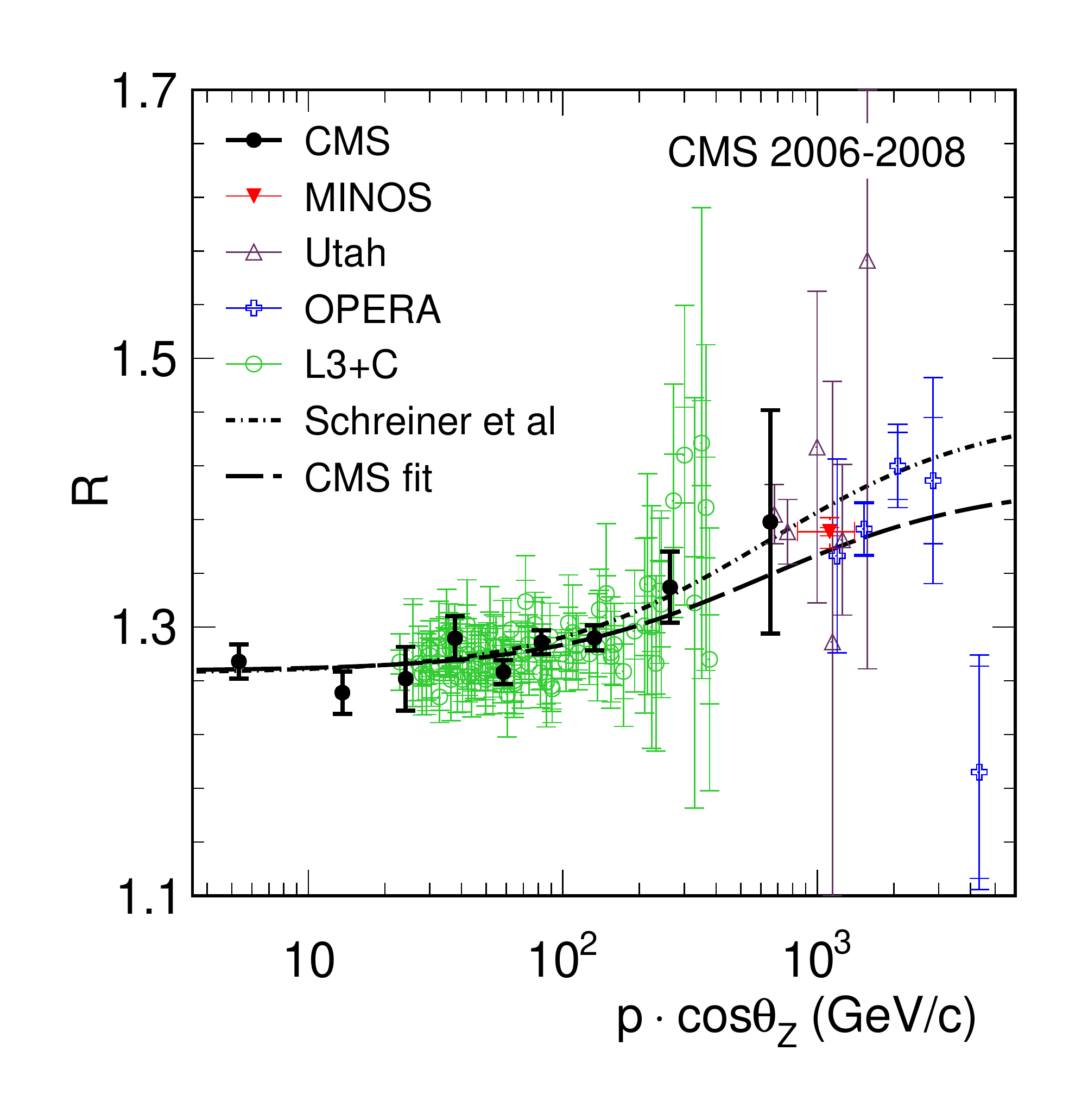}}
   \caption{(a) The three CMS results, and their combination, as a function
            of the muon momentum. Data points are placed at the bin average,
            with the points from the standalone and global-muon analyses offset
            horizontally by $\pm$10\% for clarity.
            (b) The CMS result, as a function of the vertical
            component of the muon momentum, together with some previous measurements
            and a fit of the pion-kaon model to the CMS data.
   \label{fig:results}}
\end{figure*}

The results of the three analyses are shown in Fig.~\ref{fig:results}~(a), as 
a function of the muon momentum.
In the region where the results overlap,
agreement between them is good, so the individual 
analyses are combined using a standard prescription~\cite{BLUE}.
Within each analysis, some systematic uncertainties are assumed to be
correlated between momentum bins: trigger efficiency, momentum scale,
charge misassignment and asymmetries in the muon losses due to the
detector acceptance.
In the global and standalone-muon analyses, systematic 
uncertainties from material densities, event selection, alignment, and 
magnetic field, are mostly uncorrelated between momentum bins, and are 
treated as fully uncorrelated. On the other hand, they are correlated 
between the two analyses.

The combined data points are given in Table~\ref{tab:results} as a function of $p$
and $p\cos\theta_\mathrm{z}$. They are shown in Fig.~\ref{fig:results}~(a) as a
function of $p$,
and in Fig.~\ref{fig:results}~(b) as a function of $p\cos\theta_\mathrm{z}$.

\begin{table*}[ht]
\caption{The muon charge ratio $R$ from the combination of all three CMS analyses,
as a function of $p$ and $p\cos\theta_\mathrm{z}$, in \GeVc{}, together with the combined
statistical and systematic relative uncertainty, in \%.}
\label{tab:results}
\begin{center}
\begin{tabular}{c r c c || c r c c}
\hline
$p$ range                      & $\langle p \rangle $ & $R$ & uncertainty &
$p\cos\theta_\mathrm{z}$ range & $\langle p \cos \theta_\mathrm{z}\rangle $ & $R$ & uncertainty \\
\hline
\phoo 5 -- \pho 10 &   7.0 & 1.250 & 2.45 &  \,2.5 -- \pho 10 &   5.3\phoo & 1.274 & 0.99 \\
\pho 10 -- \pho 20 &  13.7 & 1.277 & 0.85 &\pho 10 -- \pho 20 &  13.6\phoo & 1.251 & 1.26 \\
\pho 20 -- \pho 30 &  24.2 & 1.276 & 1.34 &\pho 20 -- \pho 30 &  24.1\phoo & 1.262 & 1.88 \\
\pho 30 -- \pho 50 &  37.8 & 1.279 & 1.10 &\pho 30 -- \pho 50 &  37.7\phoo & 1.292 & 1.27 \\
\pho 50 -- \pho 70 &  58.5 & 1.275 & 0.54 &\pho 50 -- \pho 70 &  58.4\phoo & 1.267 & 0.71 \\
\pho 70 --     100 &  82.5 & 1.275 & 0.68 &\pho 70 --     100 &  82.4\phoo & 1.289 & 0.70 \\
    100 --     200 & 134.0 & 1.292 & 0.52 &    100 --     200 & 133.1\phoo & 1.292 & 0.72 \\
    200 --     400 & 265.8 & 1.308 & 1.29 &    200 --     400 & 264.0\phoo & 1.330 & 1.99 \\
    $>400$         & 698.0 & 1.321 & 3.98 &    $>400$         & 654.0\phoo & 1.378 & 6.04 \\
\hline
\end{tabular}
\end{center}
\end{table*}

\subsection{Charge ratio below 100 \GeVc{}}

In the region $p < 100 \GeVc{}$ there are measurements in six $p$ bins.
Three bins are covered by all three analyses, with the surface-based MTCC analysis
extending the reach to three lower-momentum bins.
These twelve data points are combined into a single value of the charge ratio using
the same prescription and scenario for correlations as for the overall combination 
described in the above section. This yields a charge ratio of
$1.2766 \pm 0.0032 \, \mathrm{(stat.)} \pm 0.0032 \, \mathrm{(syst.)}$, 
with a $\chi^2/\mathrm{ndf} = 7.3/11$, in good agreement
with previous measurements~\cite{baxendale,rastin,hebbeker,L3C}
and representing a significant improvement in precision.

Repeating this fit in the $p \cos\theta_\mathrm{z}$ region below 100~\GeVc{}
yields a charge ratio of 
$1.2772 \pm 0.0032 \, \mathrm{(stat.)} \pm 0.0036 \, \mathrm{(syst.)}$, with a $\chi^2/\mathrm{ndf} = 15.3/11$.
The higher $\chi^2/\mathrm{ndf}$ indicates that the data in this
$p \cos\theta_\mathrm{z}$ region 
have a lower probability of being consistent with a flat charge ratio. 
Fitting just the region $p \cos\theta_\mathrm{z} <70$~\GeVc{} yields a charge ratio of 
$1.2728 \pm 0.0039 \, \mathrm{(stat.)} \pm 0.0040 \, \mathrm{(syst.)}$ with a 
$\chi^2/\mathrm{ndf} = 4.0/8$, consistent with the flat charge ratio hypothesis.

\subsection{Charge ratio in the 5\GeVc{} to 1\TeVc{} momentum range}

Considering the full $p\cos\theta_\mathrm{z}$ range measured, a
rise in the charge ratio is seen, as shown in Fig.~\ref{fig:results}~(b).
Comparing to previous measurements in the same momentum ranges, the CMS results
agree well where there is overlap: with the L3+C measurement~\cite{L3C}
below 400\GeVc{}, and with the UTAH~\cite{utah}, MINOS~\cite{Adamson:2007ww} and
OPERA~\cite{OPERA} measurements above 400\GeVc{}. 
Measurements by other experiments in the range
5--20~\GeVc{}~\cite{baxendale,rastin,hebbeker,L3C,BESS2004}
are not shown in the plot; they are consistent with the
constant value fitted in the CMS data.

Models of cosmic ray showers provide an explanation for the rise in charge ratio 
at higher momentum.
Based on the quark content of protons,
and on the observation that primary cosmic ray particles are mostly positive,
the ratio $\pi^+/\pi^-$ is predicted to be around 1.27~\cite{fiorentini}.
Due to the phenomena of associated production, the charge ratio of strange
particles such as kaons is expected to be even higher.

The expected muon spectrum has been parametrized~\cite{gaisser} based on the 
interactions
of primary cosmic ray particles and on the decays of secondary particles, and from this
parametrization, the charge ratio can be extracted~\cite{arXiv09063726}
as a function of the fractions of all pion and kaon decays that yield positive muons,
$f_\pi$ and $f_K$, respectively.
These constants are not known {\sl a priori}, and must be inferred from data.

A fit performed to the combined CMS charge ratio measurement 
in the entire $p\cos\theta_\mathrm{z}$ region,
with a fixed relative amount of kaon production~\cite{gaisser},
yields $f_\pi = 0.553 \pm 0.005$, 
and \mbox{$f_K = 0.66 \pm 0.06$}, with a $\chi^2/\mathrm{ndf} = 7.8/7$.
Figure~\ref{fig:results}~(b) shows the fit to CMS data only, together with a
fit performed on some previous measurements by L3+C and MINOS~\cite{arXiv09063726}.

\section{Conclusions}

We have measured the flux ratio of positive- to negative-charge cosmic ray
muons, as a function of the muon momentum and its vertical component, using
data collected by the CMS experiment in 2006 and 2008. The result is in
agreement with previous measurements by underground experiments. This is the
most precise measurement of the charge ratio in the momentum region below
0.5\TeVc{}. It is also the first physics measurement using muons with the
complete CMS detector.

\section*{Acknowledgments}

We thank the technical and administrative staff at CERN and other CMS institutes.
This work was supported by the Austrian Federal Ministry of Science and Research;
the Belgium Fonds de la Recherche Scientifique, and Fonds voor Wetenschappelijk
Onderzoek; the Brazilian Funding Agencies (CNPq, CAPES, FAPERJ, and FAPESP);
the Bulgarian Ministry of Education and Science; CERN; the Chinese Academy of Sciences,
Ministry of Science and Technology, and National Natural Science Foundation of China;
the Colombian Funding Agency (COLCIENCIAS); the Croatian Ministry of Science, Education
and Sport; the Research Promotion Foundation, Cyprus; the Estonian Academy of
Sciences and NICPB; the Academy of Finland, Finnish Ministry of Education, and
Helsinki Institute of Physics; the Institut National de Physique Nucl\'eaire
et de Physique des Particules~/~CNRS, and Commissariat \`a l'\'Energie Atomique,
France; the Bundesministerium f\"ur Bildung und Forschung, Deutsche
Forschungsgemeinschaft, and Helmholtz-Gemeinschaft Deutscher Forschungszentren, Germany;
the General Secretariat for Research and Technology, Greece; the National Scientific
Research Foundation, and National Office for Research and Technology, Hungary;
the Department of Atomic Energy, and Department of Science and Technology, India;
the Institute for Studies in Theoretical Physics and Mathematics, Iran; the Science
Foundation, Ireland; the Istituto Nazionale di Fisica Nucleare, Italy; the Korean
Ministry of Education, Science and Technology and the World Class University program
of NRF, Korea; the Lithuanian Academy of Sciences; the Mexican Funding Agencies
(CINVESTAV, CONACYT, SEP, and UASLP-FAI); the Pakistan Atomic Energy Commission;
the State Commission for Scientific Research, Poland; the Funda\c{c}\~ao para a
Ci\^encia e a Tecnologia, Portugal; JINR (Armenia, Belarus, Georgia, Ukraine, Uzbekistan);
the Ministry of Science and Technologies of the Russian Federation, and Russian
Ministry of Atomic Energy; the Ministry of Science and Technological Development
of Serbia; the Ministerio de Ciencia e Innovaci\'on, and Programa Consolider-Ingenio 2010,
Spain; the Swiss Funding Agencies (ETH Board, ETH Zurich, PSI, SNF, UniZH, Canton
Zurich, and SER); the National Science Council, Taipei; the Scientific and Technical
Research Council of Turkey, and Turkish Atomic Energy Authority; the Science and
Technology Facilities Council, UK; the US Department of Energy,
and the US National Science Foundation.

Individuals have received support from the Marie-Curie IEF program (European Union);
the Leventis Foundation; the A. P. Sloan Foundation; the Alexander von Humboldt Foundation;
and the Associazione per lo Sviluppo Scientifico e Tecnologico del Piemonte (Italy).

\bibliography{auto_generated}

\cleardoublepage\appendix\section{The CMS Collaboration \label{app:collab}}\begin{sloppypar}\hyphenpenalty=5000\widowpenalty=500\clubpenalty=5000\textbf{Yerevan Physics Institute,  Yerevan,  Armenia}\\*[0pt]
V.~Khachatryan, A.M.~Sirunyan, A.~Tumasyan
\vskip\cmsinstskip
\textbf{Institut f\"{u}r Hochenergiephysik der OeAW,  Wien,  Austria}\\*[0pt]
W.~Adam, T.~Bergauer, M.~Dragicevic, J.~Er\"{o}, C.~Fabjan, M.~Friedl, R.~Fr\"{u}hwirth, V.M.~Ghete, J.~Hammer\cmsAuthorMark{1}, S.~H\"{a}nsel, M.~Hoch, N.~H\"{o}rmann, J.~Hrubec, M.~Jeitler, G.~Kasieczka, W.~Kiesenhofer, M.~Krammer, D.~Liko, I.~Mikulec, M.~Pernicka, H.~Rohringer, R.~Sch\"{o}fbeck, J.~Strauss, A.~Taurok, F.~Teischinger, W.~Waltenberger, G.~Walzel, E.~Widl, C.-E.~Wulz
% ...
\vskip\cmsinstskip
% ...
1:~~Also at CERN, European Organization for Nuclear Research, Geneva, Switzerland\\
% ...

\end{sloppypar}
\end{document}